\documentclass{osa-article}

\journal{oe}
 
\newcommand*\moire{moir\'{e} }
\newcommand*\Moire{Moir\'{e} }

\usepackage{tabularx}

\begin{document}

\title{Demonstration of focal length tuning by rotational varifocal \moire metalens in an ir-A wavelength}

\author{Kentaro Iwami,\authormark{1,*} Chikara Ogawa,\authormark{1} Tomoyasu Nagase\authormark{1}, and Satoshi Ikezawa\authormark{1}}

\address{\authormark{1}Department of Mechanical Systems Engineering, Tokyo University of Agriculture and Technology, Koganei, Tokyo 184--8588 Japan}

\email{\authormark{*}k\_iwami@cc.tuat.ac.jp} 

\homepage{https://nmems.lab.tuat.ac.jp/en/} 


\begin{abstract*}
This paper reports an experimental demonstration of \moire metalens which shows wide focal length tunability from negative to positive by mutual angle rotation at the wavelength of 900 nm. The \moire metalens was developed using high index contrast transmitarray meta-atoms made of amorphous silicon octagonal pillars, which is designed to have polarization insensitivity and full 2$\pi$ phase coverage. The fabricated \moire metalens showed focal length tunability at the ranges between $\pm$1.73 -- $\pm$5 mm, which corresponds to the optical power ranges between $\pm$578 -- $\pm$200 m$^{-1}$ at the mutual rotation between $\pm 90$ degrees. 
\end{abstract*}

\section{Introduction}
Metasurfaces, a planar branch of metamaterials, have attracted significant attention because of its possibility to tailor optical wavefront by arranging subwavelength patterns (meta-atoms) on the surface.\cite{Zheludev2012,Glybovski2016,Yu2014}
They are able to show unique optical properties that cannot be achieved in natural materials and have a high affinity for micro/nanofabrication methods including lithography, deposition, and etching. 
Therefore, metasurfaces have opened up many research fields and applications including lens\cite{Ni2013,Khorasaninejad2016,Wang2018,Suzuki2020}, retarders and waveplates\cite{Kats2012,Yu2012,Ishii2015,Ishii2016}, vector beam converters\cite{Genevet2012,Iwami2012,Hakobyan2016}, color filters\cite{Ikeda2012,Nagasaki2018a,Nagasaki2018}, holography\cite{Huang2013,Yifat2014,Huang2015,Wan2016,Izumi2020}, zero refractive index materials\cite{Suzuki2020a}, and others. 
It should be emphasized that mechanical deformation or change of mutual geometric position of metasurfaces offers novel functionality and tunability for their optical properties. 
Based on this idea, reconfigurable metasurfaces have been studied including tunable transmittance\cite{Ou2011}, color tuning\cite{Honma2012}, active phase shifters\cite{Yamaguchi2016, Shimura2018}, and so on. 

Metasurface lens, or metalens, has been attracted many interests because of its thinness and lightweight.
Especially, recent developments of lossless dielectric metalenses\cite{Arbabi2015,Khorasaninejad2016,Wang2018} have been boosted attention to this field. The integration of the metalens with microelectromechanical systems (MEMS) enables both compactness and functionalities\cite{Roy2018}. 
Tunable focal length, or varifocal lens, is one of the most promising and expected functionalities of metalens.
Similar to conventional refractive lens doublet, the longitudinal motion of lenses along with the optical axis based on a MEMS actuator has been demonstrated\cite{Arbabi2018}. 
However, the pull-in instability effect of electrostatic parallel plates actuator limits the travel range of metalens\cite{Hung1999,Nemirovsky2005} and resulting in a narrow tunable range of focal length. 

As metalens has much design freedom, a tuning method of the focal length is not necessary to be similar to conventional refractive lenses.
Actually, studying diffractive optical elements (DOEs) or diffractive lenses may bring many advantages to metalens.
In fact, DOEs-inspired tunable metalens, Alvarez metalens, has been experimentally demonstrated, and focal length tuning based on lateral movement of a pair of lenses was achieved\cite{Zhan2017,Colburn2018a,Colburn2019}. Furthermore, the integration with the MEMS actuator to the Alvarez lens dynamic focal length control with the compact actuation mechanism\cite{Han2020a}.
However, the experimental tuning range of the focal length was limited to the positive region, in contrast to the potential capability to both positive and negative of the Alvarez configuration. Furthermore, the lateral motion of the Alvarez lens limits the effective area (the aperture) of the lens.

\Moire lens is a doublet of axially-asymmetric lenses that show tunable focal length with mutual rotation as shown in Figure \ref{fgr:principle}(a)\cite{Bernet2008, Bernet2017}.
It can offer a wide range of focal lengths from negative to positive and has been realized using diffractive optical elements (DOEs)\cite{Bernet2013,Heide2016}.
\Moire metalens has also been studied with numerical simulations\cite{Yilmaz2019, Liu2019}, and quite recently experimentally demonstrated in microwave frequency\cite{Guo2019} and the infrared-B (ir-B) wavelength\cite{Wei2020}. However, the continuous promotion of shorter wavelengths and larger diameters are required increasingly on metalenses, despite the manufacturing difficulty\cite{Park2019a,She2018}, especially in the axially-asymmetric lenses. 

Here, we experimentally demonstrate tunable focal length with \moire metalens at the ir-A wavelength of 900 nm using polarization-insensitive meta-atom based on high index contrast transmitarrays (HCTAs), specifically made of amorphous silicon (a-Si) in this case.
A-Si octagonal pillars were designed to have polarization insensitivity and full 2$\pi$ phase coverage at the wavelength of 900 nm in order to use a character projection (CP) type electron beam lithography (EBL) apparatus for high-throughput drawing.
The metalens is designed to satisfy the phase distribution of \moire metalens as detailed below by mapping corresponding pillars to each lattice point with a hexagonal basis.
The designed metalens was successfully fabricated with a diameter of 2 mm, which is the largest among EBL-drawn varifocal metalenses on a glass substrate using simple a-Si sputter deposition, EBL using CP, metal mask lift-off, and reactive ion etching (RIE) of silicon.
The fabricated metalens showed focal length tunability at the ranges between $\pm$ 1.73 -- $\pm$ 5 mm at the mutual rotation between $\pm 90$ degrees at the wavelength of 900 nm. 
The results reported here demonstrate a proof of concept for the focal length tuning with a mutual rotation of lens components based on \moire lens configuration at the wavelength, and pave the way to applications in optical frequencies including near-infrared, visible, and ultraviolet wavelengths.

\section{Design and fabrication of \moire metalens}

\begin{figure}
	\centering\includegraphics[width=13cm]{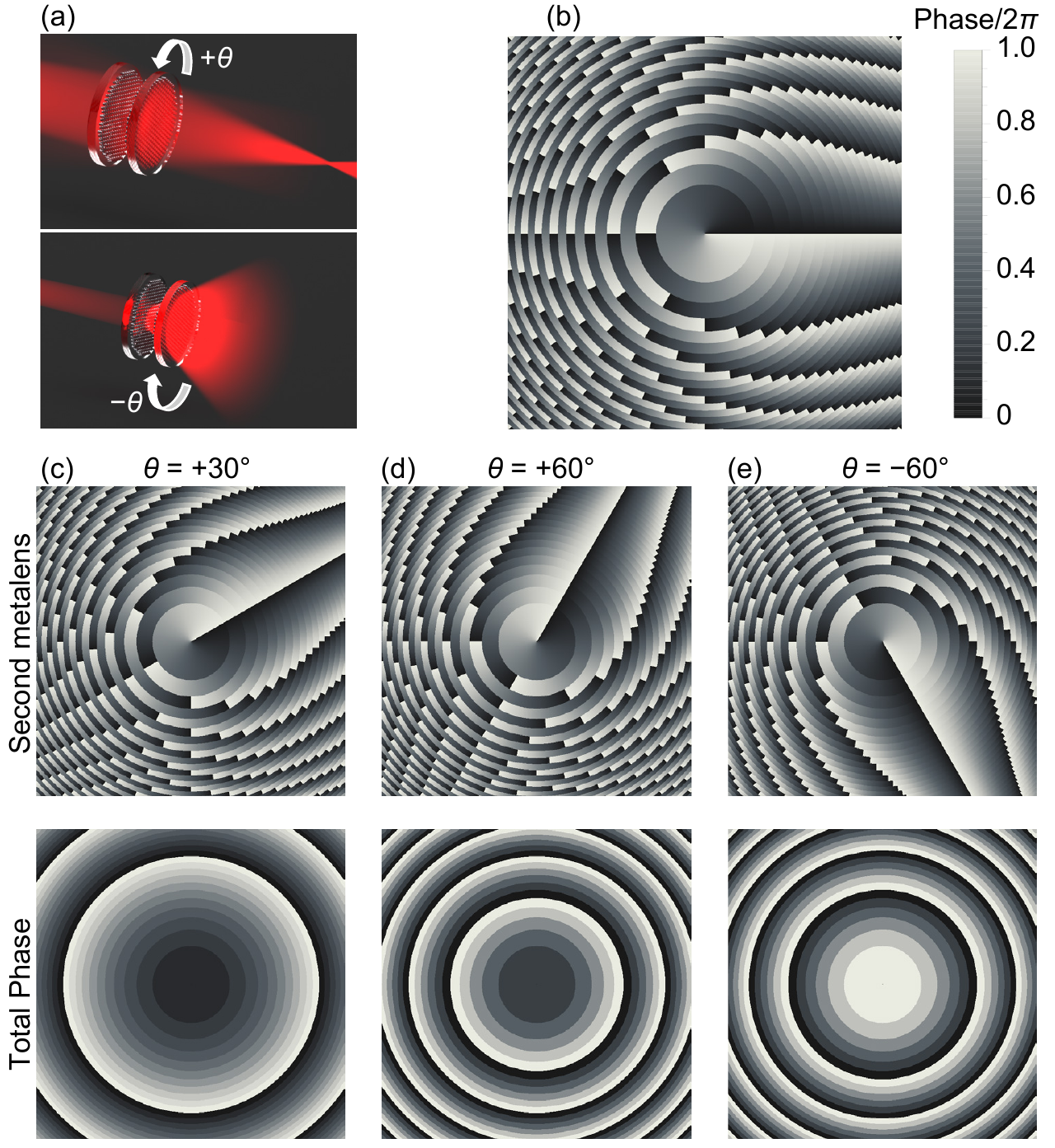}
	\caption{Principle of \Moire metalens. Schematic drawing of working principle (a). Phase distribution of one of two lenses (b). Phase distributions of second lens and combined \moire metalens for various mutual rotation angles $\theta$ for focal length $f$ tuning (c-e). Note that the same color legends are used through (b-e).}
	\label{fgr:principle}
\end{figure}

Figure\ref{fgr:principle} shows a schematic drawing of \moire metalens.
As shown in Fig. \ref{fgr:principle}(a), \moire metalens consists of two metalenses, and their total focal length can be tuned by mutual rotation from negative to positive.
Each part of a \moire metalens pair were designed to have transmission function distributions in polar coordinates $(r, \varphi)$\cite{Bernet2008},
\begin{eqnarray}
	T_1 (r,\varphi) &=& \exp \left\lbrace i \textrm{round}(a r^2) \varphi \right\rbrace , and \nonumber \\
	T_2 (r,\varphi) &=& \exp \left\lbrace -i \textrm{round}(a r^2) \varphi \right\rbrace ,
	\label{eqn:moireT}
\end{eqnarray}
where $a$ is a constant. 
The phase distribution of the first lens $T_1$ is shown in Fig. \ref{fgr:principle}(b).
With mutual rotation $\theta$, the joint transmission function can be given by
\begin{equation}
	T_\textrm{joint}=T_1(r, \varphi) T_2(r, \varphi-\theta)=\exp \left\lbrace i \textrm{round} (a r^2)\theta \right\rbrace.
	\label{eqn:jointT}
\end{equation}

This equation is similar to that of a spherical lens under paraxial approximation $T=\exp (i \pi r^2/f \lambda)$, where $f$ is the focal length and $\lambda$ is the wavelength.
So, by adjusting the constant $a$ to satisfy the equation $f^{-1}=a \theta \lambda/\pi$, the optical power $f^{-1}$ becomes proportional to the mutual rotation angle $\theta$. 
Note that the round function is used to avoid the sectoring effect\cite{Bernet2008}.

Figures \ref{fgr:principle}(c-e) show phase distributions of the second metalens with mutual rotation angles of +30, +60, and --60 degrees, respectively, together with total phase distributions superimposed with the first metalens, shown in Fig. \ref{fgr:principle}(b). As shown in Figs. \ref{fgr:principle}(c) and (d), positive rotation angles create total phase distribution similar to convex Fresnel lens, and a larger rotation angle makes a higher phase gradient, which corresponds to short focal length and higher optical power.
In contrast, as shown in Fig. \ref{fgr:principle}(e), a negative rotation angle creates concave-like phase distribution.
As shown here, by adopting \moire metalens, a wide tuning range of focal length from negative to positive can be expected.

\begin{figure}
	\centering\includegraphics{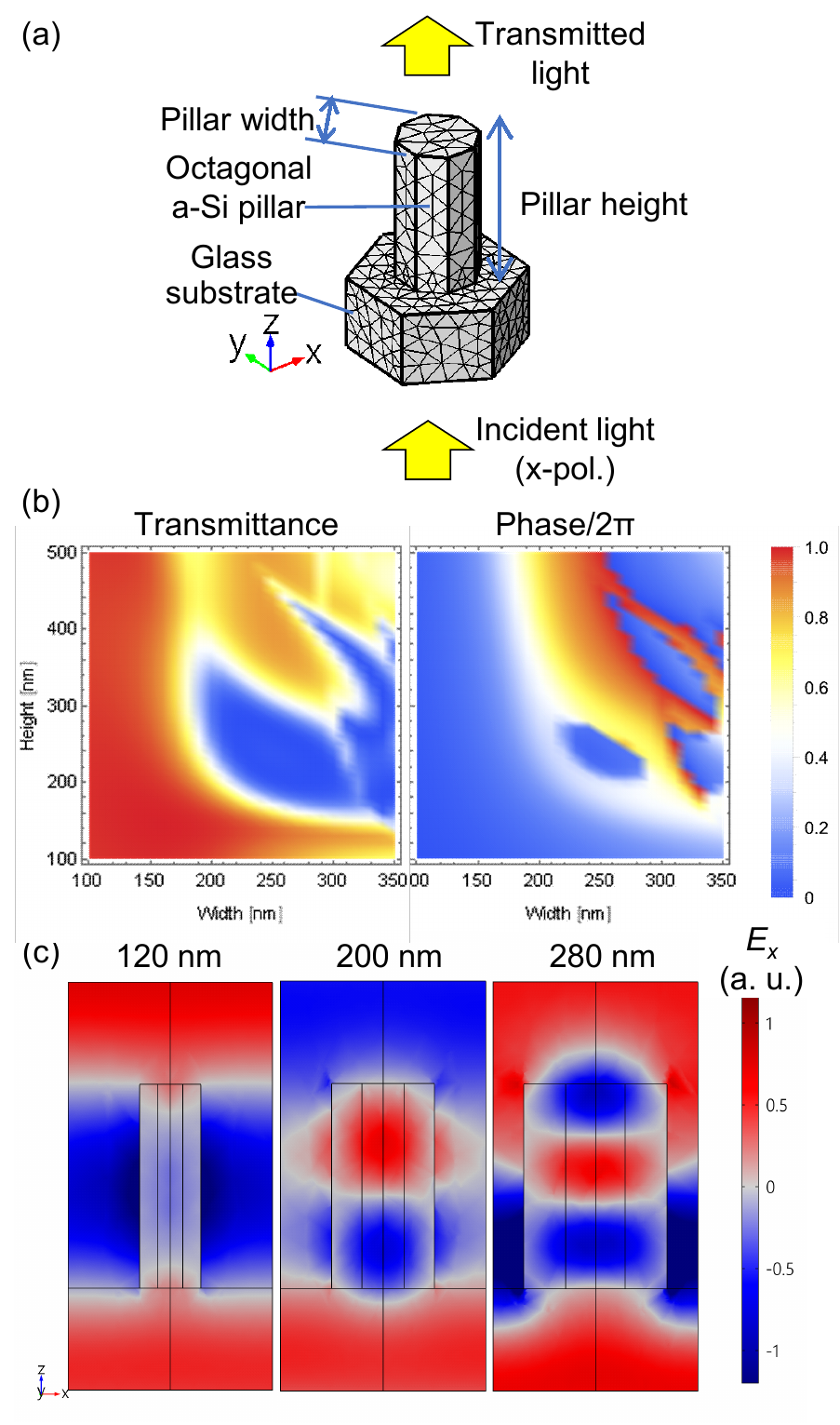}
	\caption{COMSOL simulation of a-Si octagonal pillar meta-atom. Calculation model (a). Parameter mapping of transmittance and phase/2$\pi$, sweeping pillar height and width (b). Electric field ($E_x$) distribution of a-Si octagonal pillars with the height of 400 nm and the widths of 120 nm, 200 nm, and 280 nm, respectively (c).}
	\label{fgr:COMSOL}
\end{figure}

An a-Si octagonal pillar HCTAs have been adopted as polarization insensitive meta-atoms\cite{Arbabi2015}. An octagonal shape was chosen in order to use CP of the EBL apparatus for the high-throughput fabrication as detailed later. An electromagnetic simulation was performed using a commercially available finite element method software COMSOL Multiphysics ver. 5.1 (COMSOL Inc., USA) as shown in Fig. \ref{fgr:COMSOL}. Figure \ref{fgr:COMSOL}(a) shows a schematic of the simulation setup. An octagonal pillar was situated on the silica glass substrate with a period of 400 nm. The surrounding medium of the pillar is air (not shown). The $x$-polarized light at the wavelength of 900 nm incidents from the glass bottom side, then transmittance and phase delay of the pillar was evaluated at the output port (the air side). Froquet periodic boundary conditions are applied to boundaries parallel to $z$-axis. An a-Si material parameter was used to the Si pillar\cite{Pierce1972}.

Fig. \ref{fgr:COMSOL}(b) shows the parameters sweep results of transmittance and phase/2$\pi$, by changing both pillar width and height with 10 nm steps. As a result, it was found that the region around height$> 380$ nm is able to achieve both high transmittance and full 2$\pi$ phase coverage. Fig. \ref{fgr:COMSOL}(c) shows the $x-$component of electric field $E_x$ distribution at the pillar height of 400 nm and widths of 120, 200, 280 nm, respectively. It is clearly shown that phase difference between 200, 280 nm pillars, and 120 nm pillar at the output port (Top of the image) are corresponding to $\pi$, and $2\pi$, respectively. Therefore, considering the fabrication process limitation for high aspect ratio structure, we adopted 400-nm-height pillars as the meta-atom.

Using these octagonal pillars as meta-atoms, \moire metalens with a diameter of 2 mm was designed. The "a" parameter was set to 0.001283 rad$^{-1}$\textmu m$^{-2}$ to have maximum optical powers of $f^{-1}=\pm (1.73  \textrm{ mm})^{-1}$ at the mutual rotation angles $\theta=\pm 90$ degrees. It corresponds to the maximum numerical aperture (NA) of the \moire metalens to be 0.5. A GDSII CAD file of the \moire metalens was prepared by mapping pillars with corresponding phase delay satisfying the phase distribution described by $T_1$ of the equation \ref{eqn:moireT} to each lattice point with a hexagonal basis with the pitch of 400 nm using a Python library gdspy\cite{Gabrielli2020}.

A metalens was fabricated on a quartz glass substrate. The detailed fabrication process is available in Supplement 1 Section S1. An amorphous silicon film with a thickness of 400 nm was sputter-deposited. The optical properties of the deposited silicon thin film are shown in Supplement 1 Section S2. A lens pattern with a diameter of 2 mm was drawn by electron beam lithography with octagonal CP (Advantest F7000-S, Advantest, Japan) using dedicated octagonal stencil masks in order to achieve high drawing throughput. The GDSII CAD file described above was converted into CP shot data with proximity effect compensation (PEC). Octagonal CPs with the nominal width-across-flats from 120 to 280 nm with a step of 10 nm were used. Thanks to the high-throughput of CPs, a 2-cm-diameter \moire metalens, which consists of approximately $1.7\times 10^7$ meta-atoms with the GDSII file size of 1.2 GB, can be drawn less than 5 minutes exposure. After the resist development, an aluminum mask was patterned on the a-Si film through the lift-off process. Si pillars are formed by reactive ion etching. Finally, the aluminum mask was removed by a wet chemical. 

\section{Results and discussion}

\begin{figure}
	\centering\includegraphics{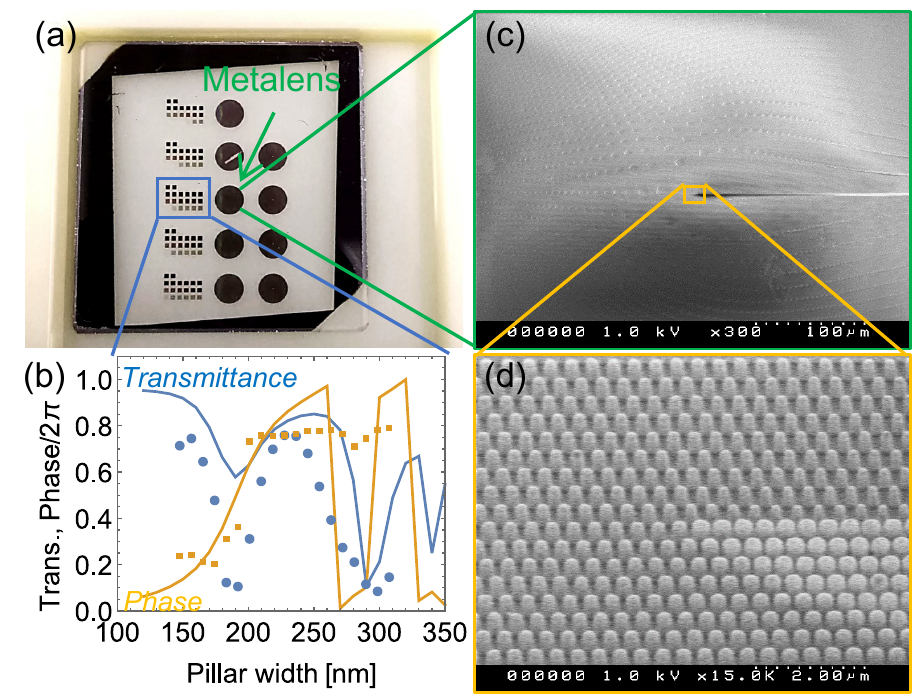}
	\caption{Fabricated metalens. A photograph of fabricated lenses on a 2-cm-square glass substrate (a). Nine circles are 2-mm-diameter metalenses. Blue-indicated square patterns are pillar arrays with constant widths for transmission property measurements. Transmission properties at the wavelength of 900 nm(b). Dots indicate measurement results, and lines indicate COMSOL simulation. Blue and orange correspond to transmittance and phase/2$\pi$, respectively. SEM image of the lens (c) and its close-up to the center (d), observed with 45$^\circ$ tilting angles. Width-distributed pillars are successfully fabricated.}
	\label{fgr:SEM}
\end{figure}

Figure \ref{fgr:SEM} shows a photograph of fabricated lenses on a 2-cm-square glass substrate (a), transmission properties (b), and scanning electron microscope (SEM) images (c-d). In Fig. \ref{fgr:SEM}(a) Nine circles are 2-mm-diameter metalenses. Blue-indicated square patterns are pillar arrays with constant widths for transmission property measurements. Fig. \ref{fgr:SEM}(b) shows the transmission properties of the pillars as a function of pillar width at the wavelength of 900 nm. Dots indicate measurement results, and lines indicate COMSOL simulation. Blue and orange correspond to transmittance and phase/2$\pi$, respectively. Data points in Fig. \ref{fgr:SEM}(b) is based on measured widths of pillars as detailed in the Supplement 1 Section S3.
The phase delays of the pillars were measured by the interference method\cite{Miyata2019a}, using a laser diode and microspectroscopy system (Techno-Synergy, DF-1037) detailed in the Supplement 1 Section S4.
Fig. \ref{fgr:SEM}(c) shows the SEM image of the lens, and \ref{fgr:SEM}(d) shows its close-up to the center, observed with 45$^\circ$ tilting angles. Width-distributed pillars are successfully fabricated. Because of fabrication limitation, pillars become not octagonal but cylindrical.

As shown in Fig. \ref{fgr:SEM}(b), as the width increases, phase delay rapidly increases on both simulation and experiment. Although the experimental transmittance becomes smaller than that of simulation, the position of dips is similar. 
The difference between experiment and simulation is considered to be attributed to fabrication error, including the rounded shape of the pillar top as shown in \ref{fgr:SEM}(d).

\begin{figure}
	\centering\includegraphics[width=13cm]{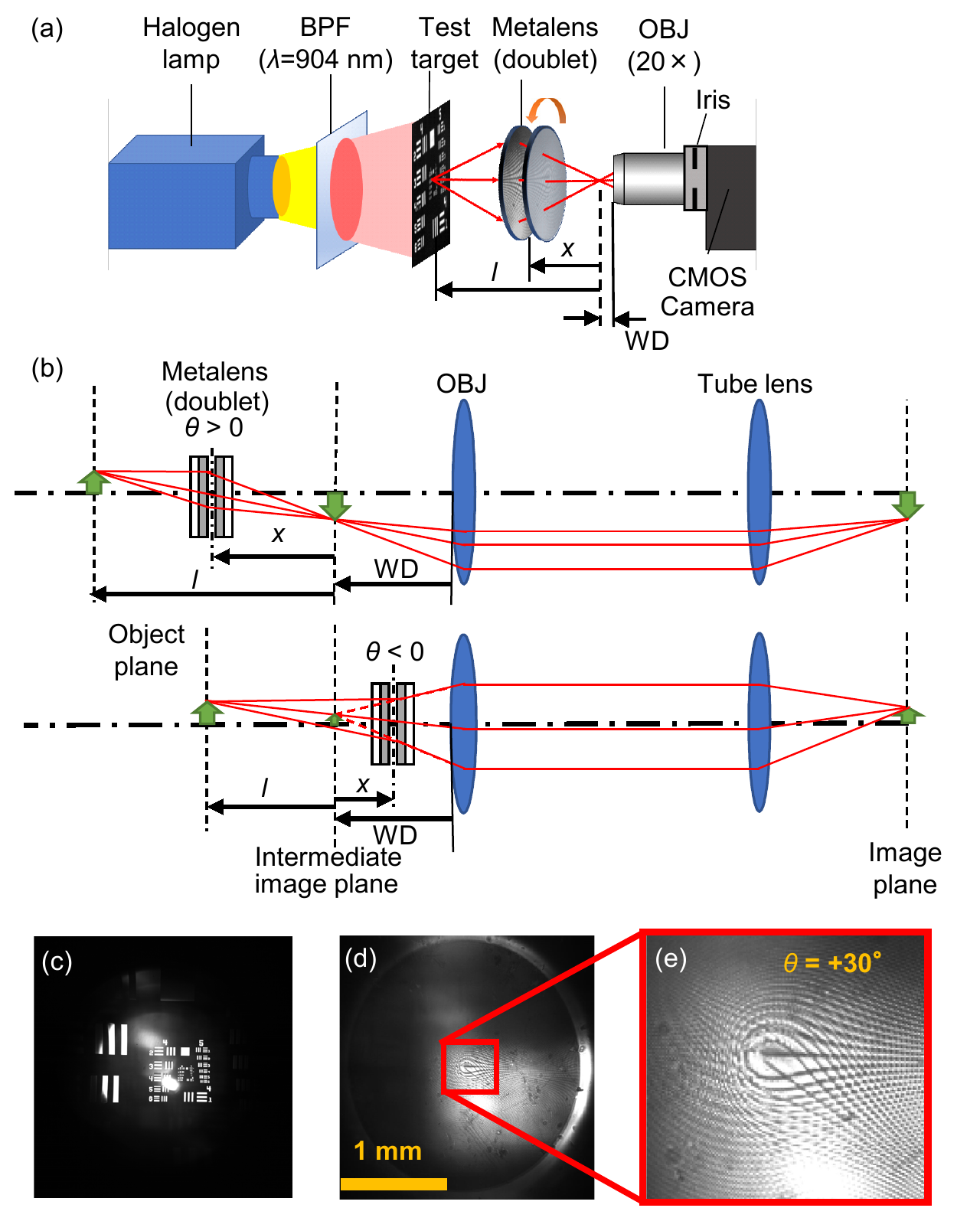}
	\caption{Experimental setup for focal length measurement. Optical system (a). Detailed ray illustrations for both positive (upper) and negative (lower) rotation angles $\theta$ (b). Image focused on the target without \moire metalens (c). The iris circle can be seen. Image focused on to the \moire metalens (d) and its magnified image (e). The mutual rotation angle can be determined from the image.}
	\label{fgr:setup}
\end{figure}

Figure \ref{fgr:setup} shows the experimental setup for focal length measurement. 
The optical system is shown in Fig. \ref{fgr:setup}(a). 
A tungsten halogen white light source was used. 
And then the incident light passes through a band pass filter (\#65-184, Edmund Optics USA, OD = 4, center wavelength = 904 nm, FWHM = 10 nm) and a test target (USAF 1951).
A \moire metalens sample is prepared by assembling two metalens pieces fabricated through the above-mentioned process.
The latter lens was flipped upside down along with the horizontal axis, to satisfy the phase distribution of $T_2$ in the equation \ref{eqn:moireT}.
These lenses are aligned carefully under the microscope observation and adhered to using a tape with certain mutual rotation angles $\theta$.
The transmitted image of the USAF target was captured using a CMOS camera (DCC1545M, Thorlab Co., USA) with a 20X objective lens and an iris.
The metalens and the USAF target were mounted on micrometer stages, respectively, and their position $x$ and $l$ were measured with 0.1 mm step.

Fig. \ref{fgr:setup}(b) shows a detailed ray illustration for both positive (upper) and negative (lower) rotation angles.

Fig. \ref{fgr:setup}(c) shows a captured image without \moire metalens and focused on the test target.
The bright part which is due to the light source can be seen in the center. 
Figs. \ref{fgr:setup}(d) shows the image focused onto the \moire metalens doublet.
The lens diameter of 2 mm was almost equivalent to the iris opening.
And Fig. \ref{fgr:setup}(e) shows the magnified image of (d). 
The mutual rotation angle of $\theta=$ 30 deg. are clearly seen.
The lateral misalignment between two lenses was less than 4 \textmu m, which corresponds to 10 pixels of meta-atom.
It is considered that this misalignment does not much affect the focal length and efficiency, compared with literature on diffractive \moire lenses\cite{Bernet2008,Bernet2013}.

Using the metalens position $x$ from the image plane and target position $l$ as shown in \ref{fgr:setup}(a) under the certain mutual rotation angle $\theta$, the focal length $f(\theta)$ of the \moire metalens was calculated using the following equation,
\begin{equation}
\frac{1}{f(\theta)}=\frac{1}{l-x}+\frac{1}{x}.
\end{equation}

\begin{figure}
	\centering\includegraphics[width=12cm]{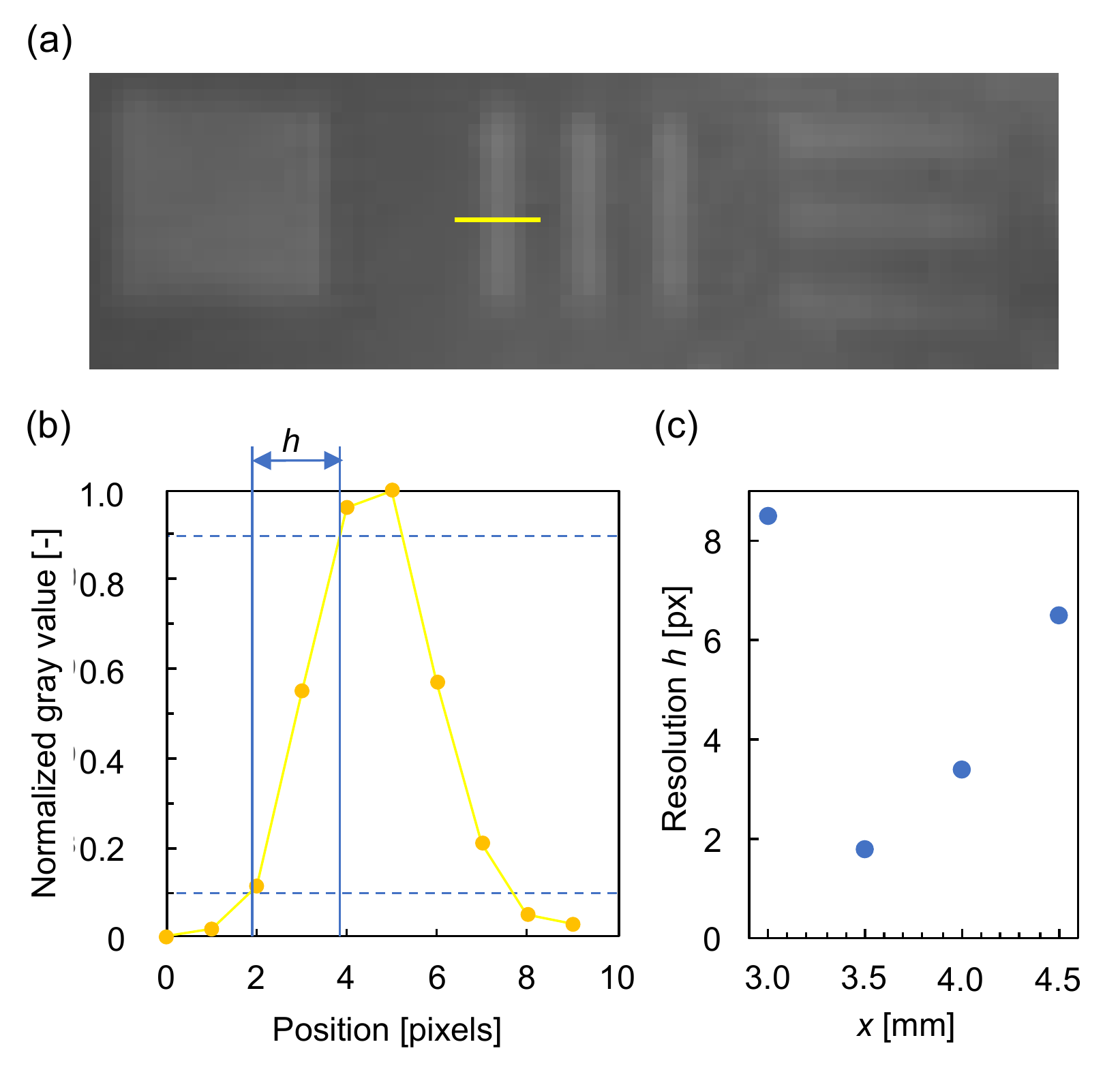}
	\caption{Determination procedure for the focal length. (a) Close-up to the group 4, number 2 of the USAF target which is obtained with the rotation angle $\theta = +60^\circ$ and $x = 3.5$ mm. The yellow line indicates the position of the normalized gray level profile shown in (b). In (b), the resolution was defined as the average of the widths between the gray values of 0.1 and 0.9. (c) Width as the function of lens position $x$. In this case the focal position was determined to $x=3.5$ mm, which corresponds to the focal length $f(60^\circ)=2.2$ mm.} 
	\label{fgr:foclen}
\end{figure}

Determination procedure of the focal length $f(\theta)$ was summarized in Fig. \ref{fgr:foclen}. Fig. \ref{fgr:foclen}(a) shows a captured image of the group 4, number 2 of the USAF target obtained with the rotation angle $\theta = +60^\circ$ and the lens position $x = 3.5$ mm. The normalized gray level profile along the yellow line is shown in  Fig. \ref{fgr:foclen}(b). In this profile, the resolution was defined as the average of the widths between the gray values of 0.1 and 0.9 with the linear interpolation between the data points.  Fig. \ref{fgr:foclen}(c) summarizes the resolution as the function of the lens position $x$. In this case, the focal position was determined to $x=3.5$ mm, which corresponds to the focal length $f(60^\circ)=2.2$ mm.

\begin{figure}
	\centering\includegraphics{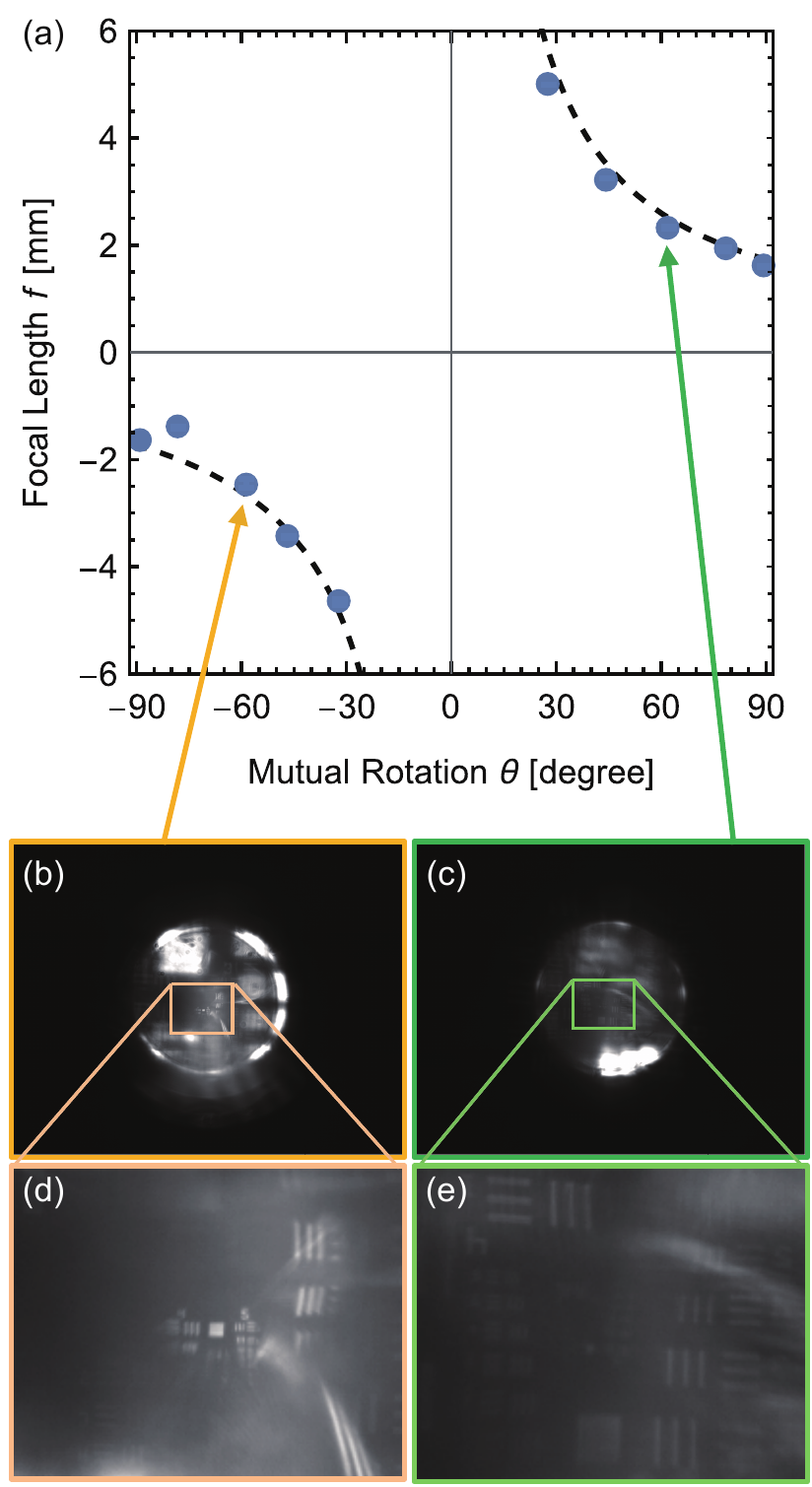}
	\caption{Measurement results of focal length (a) and typical imaging results (b-e). In (a), circle markers show experimental results obtained from the setup shown in Fig. \ref{fgr:setup}. The dashed curve shows designed characteristics $f=\pi/ (a \lambda \theta )$. (b) and (c) show captured images at the nominal rotation angles of --60 and +60 degrees, respectively, and (d) and (e) show their magnified images. Note that inverted real images are shown in (c) and (e), whereas straight virtual images are shown in (b) and (d).}
	\label{fgr:Result}
\end{figure}



Figure \ref{fgr:Result} shows the measurement results of focal length (a) and typical imaging results (b-e). In Fig. \ref{fgr:Result}(a), circle markers show experimental results obtained from the setup shown in Fig. \ref{fgr:setup}. The dashed curve shows designed characteristics $f=\pi/ (a \lambda \theta )$. Note that all standard deviations of the data points are smaller than the diameter of each marker. The experimental results agree very well with the design, and focal length tuning from negative to positive has been demonstrated at the wavelength of 900 nm. (b) and (c) show captured images at the nominal rotation angles of --60 and +60 degrees, respectively. (d) and (e) show magnified images of (b) and (c). Note that inverted real images are shown in (c) and (e), which corresponds to a positive focal length. In contrast, straight virtual images are shown in (b) and (d), which corresponds to negative focal length. However, for both negative and positive rotation angles, the contrast of the obtained images was not so high. A possible reason for this is considered as the stray light as shown in fig. \ref{fgr:setup}(c).

The obtained result of this research was summarized in Table 1 in comparison with other papers related to mechanically tunable varifocal metalenses. As shown in this table, results from this paper show a wide tunable range of both the focal length and the optical power from both negative to positive. In short, metalens can be tuned as both convex and concave lenses. Furthermore, this lens achieved the largest size of 2 mm in diameter among EB-drawn metalenses (\cite{Arbabi2018,Zhan2017,Han2020a,Wei2020}) due to the high throughput of CP exposure. 
	
\begin{table}
	\caption{Comparison of reported mechanically tunable varifocal metalens. "Size" indicates the diameter for lateral and \moire lenses, and aperture length for Alvarez lenses, respectively.}
	\label{tbl:varifocal}
	\begin{tabular}{p{7.3em}p{3.5em}p{6.4em}p{6em}p{4.8em}p{2.5em}}
		\hline
		Reference  & Method & Focal Length  [mm] & Optical Power [m$^-1$] & Wavelength [nm] & Size [mm] \\
		\hline
		Arbabi \textit{et al.} \cite{Arbabi2018} & Lateral, MEMS & 0.565 -- 0.629 & 1590 -- 1770 & 915 & 0.5  \\
 		Zhang \textit{et al.} \cite{Zhan2017}   & Alvarez        & 0.5 -- 3 & 333 -- 2000 & 633 &0.15 \\
 		Colburn  \textit{et al.} \cite{Colburn2018a} & Alvarez    & 34 -- 99 & 10 -- 29 & 1550 &10 \\
 		Colburn  \textit{et al.} \cite{Colburn2018a} & Alvarez    & 86 -- 303& 3 -- 12 & 633 &10 \\
 		Han \textit{et al.} \cite{Han2020a} &	Alvarez, MEMS & +0.18 -- +0.25&  +4000 -- +5460 &1550 & 0.5 \\
 		Wei \textit{et al.} \cite{Wei2020} & \Moire &  $\pm$3 -- $\pm$54   & $\pm$19  -- $\pm$333 & 1550 &1 \\
		This work                                    & \Moire     & $\pm$1.73 -- $\pm$5 & $\pm$200 --$\pm$578  & 900 &2\\
		\hline
	\end{tabular}
\end{table}

Although alignment and adhesion processes were used in this paper, as MEMS-actuated rotational stages have been reported, micro-rotational varifocal metalens will be realized by proper integration with actuators. And, in this case, we have tried a 2-mm-diameter lens, it would be possible to extend it to cm-scale in principle, because of the high throughput advantage of CP-based EB lithography.

\section{Conclusion}
In conclusion, focal length tuning by a rotational varifocal metalens, or \moire metalens, has been demonstrated in this paper at the ir-A wavelength of 900 nm. The metalens was designed using polarization-insensitive a-Si HCTAs as meta-atoms. The largest diameter of 2 mm among EBL-drawn varifocal metalenses was achieved because of the help of CP-based EB lithography apparatus with high drawing throughput. The fabricated metalens has shown focal length tuning at the ranges of $\pm$1.73 -- $\pm$5  mm at the mutual rotation between $\pm 90$ degrees, which corresponding to the optical power tuning range of $\pm$ 200 -- $\pm$ 578 m$^{-1}$. 

\section*{Funding}
This work was supported by the Japan Society for the Promotion of Science (JSPS) grant-in-aid Nos. 17H02754, 	19H02103, and 19H02607.

\section*{Acknowledgments}
This research was supported by the Nanotechnology Platform site at the University of Tokyo, which is supported by the Ministry of Education, Culture, Sports, Science, and Technology (MEXT), Japan. The authors thank Prof. Yoshio Mita, and Dr. Eric Lebrasseur for their help with electron beam lithography and reactive ion etching, Prof. Kuniaki Konishi for his help with ellipsometry, and Prof. Lucas H. Gabrielli for the development of gdspy.

\section*{Disclosures}
The authors declare no conflicts of interest.

\vspace{\baselineskip}
\noindent
See Supplement 1 for supporting content.

\bibliography{references}

\end{document}